\documentclass[twocolumn,showpacs,preprintnumbers,amsmath,nofootinbib,amssymb]{revtex4-1}

\usepackage{hyperref}
\usepackage{graphicx}
\usepackage{epsfig}
\usepackage{floatrow}
\usepackage{subfig}
\usepackage{dcolumn}
\usepackage{bm}

\def\be{\begin{equation}}
\def\ee{\end{equation}}
\def\bea{\begin{eqnarray}}
\def\eea{\end{eqnarray}}
\def\NO{\nonumber}
\def\gev{\mathrm{~GeV}}

\def\md{\mathrm{d}}

\begin{document}


\title{QCD Corrections to the Color-singlet $J/\psi$ Production in Deeply Inelastic Scattering at HERA}


\author{Zhan Sun}
\affiliation{School of Science, Guizhou Minzu University, Guiyang, 500025, P. R. China.}
\author{Hong-Fei Zhang}
\email{hfzhang@ihep.ac.cn (corresponding author)}
\affiliation{Department of Physics, School of Biomedical Engineering, Third Military Medical University, Chongqing 400038, China.}%
\date{\today}

\begin{abstract}
We present the first study of the QCD corrections to the color-singlet (CS) $J/\psi$ production in deeply inelastic $ep$ scattering at HERA.
The $K$-factor ranges from $0.85$ to $2.38$ in the kinematic regions we study.
In low transverse momentum regions, the $K$-factors is even smaller, and close to 1,
which indicates good convergence of the perturbative expansion.
With the QCD corrections, the CS cross section is still below the data.
At least at QCD next-to-leading order, the color-octet mechanism is necessary to describe the data.
\end{abstract}

\maketitle

Since the discovery of the $J/\psi$ meson~\cite{Aubert:1974js, Augustin:1974xw}, the production mechanism of heavy quarkonia has been a hot issue in high energy physics.
Before the first measurement of the $J/\psi$ and $\psi(2s)$ hadroproduction carried out by the CDF Collaboration~\cite{Abe:1992ww},
the color-singlet (CS) model~\cite{Einhorn:1975ua, Ellis:1976fj, Carlson:1976cd, Chang:1979nn, Berger:1980ni, Baier:1981uk, Baier:1981zz}
was generally accepted as a natural description of the heavy quarkonia production and decay mechanism.
In 1994, the nonrelativistic QCD (NRQCD) effective theory~\cite{Bodwin:1994jh} was proposed,
which successfully filled the huge gap between the QCD leading order (LO) predictions via the CS model
and the CDF measurement of the $J/\psi$ and $\psi(2s)$ hadroproduction~\cite{Braaten:1994vv, Cho:1995vh, Cho:1995ce}.
At QCD LO in the NRQCD framework,
the dominant hadroproduction mechanism of the $J/\psi$ mesons is the gluon fragmentation into a color-octet (CO) $^3S_1$ $c\bar{c}$ pair,
which produces a $J/\psi$ by emitting soft gluons in a long-distance process.
Although the CS model also permits a gluon fragmenting into a $J/\psi$ with additional two hard gluons emitted,
according to NRQCD, this mechanism substantially underestimates the production rate.
In addition to the $J/\psi$ hadroproduction, NRQCD also worked well in many other processes,
including the $J/\psi$ production in $\gamma\gamma$ fusion~\cite{Klasen:2001cu},
the $J/\psi$ photoproduction at HERA~\cite{Butenschoen:2009zy},
the $\chi_c$ meson hadroproduction~\cite{Ma:2010vd, Jia:2014jfa}, and etc.
However, one cannot overlook the controversies NRQCD is facing,
among which the $J/\psi$ polarization puzzle is the most well known and challenging one.
Many independent studies~\cite{Butenschoen:2012px, Chao:2012iv, Gong:2012ug, Shao:2012fs, Shao:2014fca, Bodwin:2014gia, Shao:2014yta, Bodwin:2015iua}
at QCD next-to-leading order (NLO) level have been performed.
Those who achieved good description of the $J/\psi$ polarization data are generally consistent with the $^1S_0^{[8]}$ dominance picture.
However, the theoretical studies on the $\eta_c$ hadroproduction indicate that~\cite{Butenschoen:2014dra, Han:2014jya, Zhang:2014ybe},
regarding the heavy quark spin symmetry, this picture violates the recent LHCb measurement~\cite{Aaij:2014bga}.
This paradox was remedied in Reference~\cite{Sun:2015pia} which found that, even without the $^1S_0^{[8]}$ dominance picture,
the $J/\psi$ polarization data can also be understood within the NRQCD framework.

To solve these problems, a new factorization theory was proposed in Reference~\cite{Ma:2017xno}.
From another angle of view, some researchers challenge the significance of the CO contributions,
and seek the way of describing the data within the CS framework.
The $J/\psi$ production in $e^+e^-$ annihilation provides an example of success of this idea.
With the QCD and relativistic corrections~\cite{Zhang:2006ay, Ma:2008gq, Gong:2009kp, Gong:2009ng, He:2009uf},
the CS contributions almost saturate the Belle data~\cite{Pakhlov:2009nj},
while the inclusion of the CO ones will generally ruin the agreement between theory and experiment.
Looking back at the $J/\psi$ hadroproduction,
QCD NLO corrections~\cite{Campbell:2007ws} enhance the differential cross sections for the CS $J/\psi$ hadroproduction
in medium and high transverse momentum ($p_t$) regions by one to two orders of magnitude,
which reduces the discrepancy between theory and data,
at the same time, change the $J/\psi$ polarization from transverse to longitudinal~\cite{Gong:2008sn}.
Due to the lack of the complete next-to-next-to-leading order results,
one cannot yet draw definite conclusions on the significance of the CO contributions implied by the $J/\psi$ hadroproduction data
(as a review, see e.g. Reference~\cite{Brambilla:2010cs}).
Such large $K$-factors are due to the fact that at QCD LO,
both the leading power (LP) and next-to-leading power (NLP) terms vanish,
and at QCD NLO, the NLP behaviour arises;
this new behaviour enhances the cross sections significantly, especially in high $p_t$ regions.

The $J/\psi$ production in deeply inelastic scattering (DIS),
which is also called the $J/\psi$ leptoproduction,
can serve as another test of the quarkonium production mechanisms.
The CS $J/\psi$ photoproduction has been studied at $\mathcal{O}(\alpha\alpha_s^3)$ in Refs.~\cite{Kramer:1994zi, Kramer:1995nb, Artoisenet:2009xh, Chang:2009uj, Li:2009fd},
which found that with the QCD corrections the CS contributions are still below the data, especially when $p_t$ is large.
For the $J/\psi$ leptoproduction, the deflection angle of the scattered lepton is larger,
accordingly the virtuality of the incident photon, which is emitted by the incident lepton and will interact in the hadronic process,
need to be taken into account in the perturbative calculation.
We define $Q^2=-q^2$, where $q$ is the momentum of the incident photon.
Due to larger $Q^2$, the $J/\psi$ leptoproduction shows better features than the $J/\psi$ photoproduction.
The $p_t$ of the $J/\psi$ yield data measured at HERA is relatively low,
thus the ratio ($x$) of the proton momentum taken by the interacting parton might be very small for the $J/\psi$ photoproduction,
and the gluon saturation effects~\cite{Mueller:1985wy} would be important.
Larger $Q^2$ can increase the value of $x$, correspondingly suppress the gluon saturation effects.
Moreover, as $Q^2$ increases, the contributions from the resolved photons are greatly suppressed,
and the perturbative expansion in $\alpha_s$ will be better as well.
Another interesting feature of the $J/\psi$ leptoproduction is that the NLP behaviour arises at QCD LO.
Unlike the $J/\psi$ hadroproduction case, no new behaviour emerges in the $J/\psi$ production in DIS at QCD NLO,
thus, the convergence of the $\alpha_s$ expansion should be better.

However, the emergence of $Q^2$ at the same time makes the computation much more complicated.
We notice that although the HERA collaborations have published abundant data,
the one-loop level phenomenological study of the $J/\psi$ production in DIS is still lacking.
The $J/\psi$ production in DIS has been studied at QCD LO in many
papers~\cite{Duke:1980sq, Baier:1981zz, Korner:1982fm, Guillet:1987xr, Merabet:1994sm, Fleming:1997fq, Yuan:2000cn, Kniehl:2001tk},
however, as indicated in our recent work~\cite{Zhang:2017dia},
all those~\cite{Fleming:1997fq, Yuan:2000cn, Kniehl:2001tk} under the NRQCD framework are based on a formalism that will lead to wrong results
when the ranges of the $J/\psi$ $p_t$ or rapidity ($y_\psi$) are not taken to cover all their possible values.
For this reason, we provided a renewed QCD LO study of the $J/\psi$ production in Reference~\cite{Sun:2017nly}.
In this paper, we study the QCD corrections to the CS $J/\psi$ production in DIS,
following the calculation framework provided in Refs.~\cite{Zhang:2017dia, Sun:2017nly}.

\begin{figure}
  \begin{center}
  {\includegraphics*[scale=0.3]{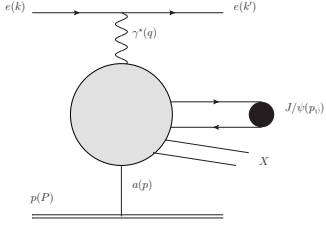}}
  \end{center}
  \caption{
Representative feynman diagram for the $J/\psi$ leptoproduction.
}
\label{fig:diag}
\end{figure}

The process for the $J/\psi$ leptoproduction is illustrated in Figure~\ref{fig:diag}.
$k$, $k'$, $P$, $p$ and $p_\psi$ are the momenta of the incident and scattered electron,
the proton, the parton generated from the proton, and the produced $J/\psi$, respectively.
The generally used invariants are defined as
\bea
&&Q^2=-q^2,~~~~W^2=(P+q)^2,~~~~z=\frac{P\cdot p_\psi}{P\cdot q}, \NO \\
&&S=2P\cdot k,~~~~s=2p\cdot q. \label{eqn:inv}
\eea
In our calculation, all the processes up to $\alpha^2\alpha_s^3$ are counted,
including $eg\rightarrow eJ/\psi g$ at both tree and one-loop level, and $ea\rightarrow eJ/\psi ij$,
in which $aij$ runs over $ggg$, $gq\bar{q}$, $qqg$, $\bar{q}\bar{q}g$, and $gc\bar{c}$.
Here we use $g$ and $q(\bar{q})$ to denote a gluon and a light quark (anti-quark), respectively.
Since all the singularities cancel in the hadronic process, namely $\gamma^\star+p\rightarrow J/\psi+X$,
where $\gamma^\star$ denotes a virtual photon,
we will directly employ the form of the leptonic tensor given in Reference~\cite{Zhang:2017dia},
which eventually leads to the same results as those by adopting the $d$-dimensional form of the leptonic tensor.
The contraction between the leptonic tensor and the hadronic one can be carried out in 4-dimensions.
Thus, the form of the leptonic tensor given in Reference~\cite{Zhang:2017dia} can be directly employed without extension to its $d$-dimensional form.
If we use the two-cutoff phase space slicing method~\cite{Harris:2001sx} to separate the divergences in the real-correction processes,
the phase space for the scattered electron and the $J/\psi$ can also be written in 4-dimensions.
Then the short-distance coefficient (SDC) for the CS $J/\psi$ production in DIS can be expressed as~\cite{Sun:2017nly}
\bea
&&\md\hat{\sigma}(e+p\rightarrow e+c\bar{c}[^3S_1^{[1]}]+X) \NO \\
&&~~=\frac{\alpha}{(4\pi)^3N_cN_sS^2}\sum_n\frac{1}{x}f_{a_n/p}(x,\mu_f)\sum_{m=1}^4C_mH^n_m \NO \\
&&~~\times\frac{\md Q^2}{Q^2}\frac{\md W^2}{W^2+Q^2}\md p_t^2\frac{\md z}{z(1-z)}\md\psi\md\Phi_X^n, \label{eqn:sdc}
\eea
where $n$ runs over all the partonic processes,
$\alpha$ is the fine structure constant, $1/(N_cN_s)$ is the spin and color average factor,
$f_{a_n/p}(x,\mu_f)$ is the parton distribution function (PDF) of a parton $a_n$ in a proton
with $x$ and $\mu_f$ being the fraction of the proton momentum taken by the parton and the factorization scale, respectively,
$p_t$ and $\psi$ are the transverse momentum of the $J/\psi$ and the azimuthal angle of the lepton plane around the $z$ axis, respectively,
and for the processes $eg\rightarrow eJ/\psi g$, $\md\Phi_X^n=\md\Phi_g=1$,
while for the real-correction processes, namely $\gamma^\star a\rightarrow J/\psi ij$,
\bea
\md\Phi_X^n&=&\md\Phi_{ij}=(2\pi)^d\delta^d(p+q-p_\psi-p_i-p_j) \NO \\
&&\times\frac{\md^{d-1}p_i}{(2\pi)^{d-1}2p_{i0}}\frac{\md^{d-1}p_j}{(2\pi)^{d-1}2p_{j0}}. \label{eqn:phixdiv}
\eea
Note that Equation~\ref{eqn:sdc} is valid in any frame of reference.
Since $\md x$ has been integrated over to eliminate the $\delta$ function which keeps the energy conservation,
the value of $x$ has been fixed in Equation~\ref{eqn:sdc} as $x=(s+Q^2)/(W^2+Q^2)$.

The expressions for $C_m$ can be found in Reference~\cite{Sun:2017nly} and are duplicated in the following as
\bea
&&C_1=A_g,~~~~C_2=\frac{4Q^2}{s^2}(A_L-2\beta A_{LT}+\beta^2A_T), \NO \\
&&C_3=\frac{4Q}{p_t^\star s}(A_{LT}-\beta A_T),~~~~C_4=\frac{1}{p_t^{\star2}}A_T, \label{eqn:c}
\eea
where
\bea
\beta=\frac{p_t^{\star2}+M^2+z^2Q^2}{2zp_t^\star Q}, \label{eqn:beta}
\eea
and
\bea
&&A_g=1+\frac{2(1-y)}{y^2}-\frac{2(1-y)}{y^2}\mathrm{cos}(2\psi^\star), \NO \\
&&A_L=1+\frac{6(1-y)}{y^2}-\frac{6(1-y)}{y^2}\mathrm{cos}(2\psi^\star), \NO \\
&&A_{LT}=\frac{2(2-y)}{y^2}\sqrt{1-y}\mathrm{cos}(\psi^\star), \NO \\
&&A_T=\frac{4(1-y)}{y^2}\mathrm{cos}(2\psi^\star). \label{eqn:a}
\eea
Hereinafter, we denote all the physical quantities in the $\gamma^\star p$ center-of-mass frame by a superscript $\star$.
$H_m$ are defined as
\bea
&&H_1=-g^{\mu\nu}H_{\mu\nu},~~~~H_2=p^\mu p^\nu H_{\mu\nu}, \NO \\
&&H_3=p^\mu p_\psi^\nu H_{\mu\nu},~~~~H_4=p_\psi^\mu p_\psi^\nu H_{\mu\nu}, \label{eqn:h}
\eea
where $H_{\mu\nu}$ is the hadronic tensor.
Note that here the long-distance matrix elements (LDMEs) have been eliminated from the hadronic tensors.
Then the cross section for the CS $J/\psi$ production can be expressed as
\bea
\md\sigma=\langle\mathcal{O}\rangle\md\hat{\sigma}(e+p\rightarrow e+c\bar{c}[^3S_1^{[1]}]+g), \label{eqn:cs}
\eea
where the CS LDME for the $J/\psi$ production, $\langle\mathcal{O}^{J/\psi}(^3S_1^{[1]})\rangle$, is abbreviated as $\langle\mathcal{O}\rangle$.

All the singularities are contained in $H^{\mu\nu}$'s and $\md\Phi_X$,
thus, we evaluate them in $d$-dimension.

We denote $H_m^{Born}$ as the $H_m$ for the process $\gamma^\star g\rightarrow c\bar{c}[^3S_1^{[1]}]g$ at tree-level,
$H_m^V$ as the virtual corrections to $H_m^{Born}$,
and $H_m^{rc}$ as the sum of all the $H_m$'s for the real-correction processes and define
\bea
&&H_m^S\equiv\int_{soft~region}\md\Phi_{ij}H_m^{rc}, \NO \\
&&H_m^{HC}\equiv\int_{hard~collinear~region}\md\Phi_{ij}H_m^{rc}. \label{eqn:hsc}
\eea
One need to be more careful in identifying the soft and hard collinear regions,
since the squared invariant mass of the incident photon is negative.
The soft region is defined in terms of the energy of the final-state gluon, $E_g$, in the $\gamma^\star g$ rest frame by $E_g\le\delta_s\sqrt{s}/2$,
while the collinear region is defined by the inequality that the inner product of two massless momenta is smaller than $\delta_cs/2$,
where $s$ is defined as in Equation~\ref{eqn:inv}.
Here, $\delta_s$ and $\delta_c$ are two arbitrary real numbers, however,
small enough to make sure that integrals of a finite function in the soft and collinear regions are negligible.
Under these definitions, most of the formulas in Reference~\cite{Harris:2001sx} remain unchanged.
To avoid double counting, the gluon-soft regions need to be excluded in the calculation of the cross sections in the hard collinear regions.
Accordingly, the integral domain of the ratio $z$ defined in Reference~\cite{Harris:2001sx} should be properly determined.
To distinguish this ratio and the inelasticity coefficient, we assign another symbol, $\xi$, to this ratio.
When $p_i$ and $p_j$ are collinear, $\xi$ is defined as $\xi=p_{i0}/(p_{i0}+p_{j0})$.
For $i=g$ and $j=q$, the integral domain of $\xi$ should be $\delta_s'<\xi<1$, where $\delta_s'=\delta_s\sqrt{s(s-Q^2)}/(s-Q^2-M^2)$,
and $M$ is the invariant mass of all the hadronic final states other than $i$ and $j$ in the partonic process.
In the process we study, $M$ is the $J/\psi$ mass, which is set to be twice of the $c$-quark mass, $m_c$, namely $M=2m_c$.
For $i,j=g$, the integral domain of $\xi$ should be $\delta_s'<\xi<1-\delta_s'$.
When $p$ and $p_i$ is collinear, the definition of $\xi$ is different, which should be $\xi=1-p_{i0}/p_0$.
The integral domain of $\xi$ should be $x<\xi<1$ for $i=q$, and $x<\xi<1-\delta_s''$ for $i=g$,
where $\delta_s''=\sqrt{(s-Q^2)/s}\delta_s$.
Here $x$ is the ratio of the parton momentum, $p-p_i$, to the incident proton momentum.
The expressions of $\delta_s'$ and $\delta_s''$ are different from those given in Reference~\cite{Harris:2001sx} due to the nonzero $Q^2$.
With the above configurations, the sum of $H_m^V$, $H_m^S$ and $H_m^{HC}$ are divergence free.
Defining
\bea
H_m^{(2)}=H_m^{Born}+H_m^V+H_m^S+H_m^{HC}, \label{eqn:H2}
\eea
one can rewrite the SDC for the CS $J/\psi$ leptoproduction as
\bea
&&\md\hat{\sigma}=\frac{\alpha}{(4\pi)^3N_cN_sS^2}\frac{\md Q^2}{Q^2}\frac{\md W^2}{W^2+Q^2}\md p_t^2\frac{\md z}{z(1-z)}\md\psi \NO \\
&&~\times\frac{1}{x}f_{a_n/p}(x,\mu_f)\sum_{m=1}^4C_m[H_m^{(2)}+\int_{H\overline{C}}\md\Phi_{ij}H_m^{rc}], \label{eqn:sdcred}
\eea
where the subscript $H\overline{C}$ means the integral is carried out in the hard noncollinear region,
and $\md\Phi_{ij}$ and $H_m^{rc}$ are evaluated in the limit $d\rightarrow4$.

To evaluate the $H_m$'s for each process,
we employ our new \textit{Mathematica} package, \textit{Malt@FDC}.
This package can automatically reduce the loop amplitudes into linear combination of master integrals,
which will be computed numerically with \textit{Looptools}~\cite{Hahn:1998yk},
and simplify the expressions of the squared amplitudes.
Our final expressions contain the $\mathcal{O}(\epsilon)$ terms of the $A_0$ and $B_0$ functions,
which is not given in the {\it Looptools} library.
These functions are computed with a new \textit{FORTRAN} package,
which will be discussed elsewhere~\cite{Zhang:b0}.
Before working on the current process,
we have applied our \textit{Malt@FDC} in tens of other processes.
All our results are consistent with those obtained by the \textit{FDC} system~\cite{Wang:2004du} and/or those given in published papers.
As an indispensable check, we studied the $Q^2\rightarrow0$ limit and compared the results with the $J/\psi$ photoproduction.
Setting $Q^2$ to be zero, replacing the leptonic tensor associated with the virtual photon propagator by $-g_{\mu\nu}$,
and implementing proper phase-space integration,
we can reproduce the $J/\psi$ photoproduction results in Refs.~\cite{Chang:2009uj, Li:2009fd}.
Another check to mention is that our results are independent of $\delta_s$ and $\delta_c$ and gauge invariant.

Abundant $J/\psi$ leptoproduction data~\cite{Adloff:1999zs, Adloff:2002ey, Aaron:2010gz, Chekanov:2005cf} have been measured at HERA.
However, most of them lie in the kinematic regions where the perturbative calculation might not be good.
In order to make the perturbation theory work better,
we constrain our concerns in the regions where $p_t^2$ and $Q^2$ are not too small and $z$ is not too large.
Adopting the selection criteria, $Q^2>12\gev^2$, $z<0.9$, and $p_t^2>6.4\gev^2$,
we find that only one set of data is availabe, which is presented in Reference~\cite{Adloff:2002ey}.
Therein, the differential cross sections with respect to $p_t^2$, $p_t^{\star2}$, $Q^2$, $W$, $y_\psi^\star$ and $z$ are measured.
To be consistent with the HERA convention,
the forward direction in the $\gamma^\star p$ rest frame are defined as that of the incident virtual photon.

Due to the $C$-parity and color conservation, $c\bar{c}(^3P_J^{[1]})$ cannot be produced at $\mathcal{O}(\alpha^2\alpha_s^2)$.
In our calculation, we completely omit the feed down contributions from $\chi_c$.
The contributions to the $J/\psi$ production from the $\psi(2S)$ feed down has been estimated in Reference~\cite{Adloff:2002ey}.
The diffractive $\psi(2S)$ are produced basically in large $z$ regions.
In the region $0.75<z<0.9$, about $6\%-20\%$ of the $J/\psi$ events come from this resource.
The cross sections of the inelastic $\psi(2S)$ production have the same behavior with that of the $J/\psi$ production.
Throughout the kinematic region we study, the contributions from the inelastic $\psi(2S)$ feed down are estimated to be about $15\%$.
The $J/\psi$ production from $b$ decay is expected in low $z$ region, and is estimated to be about $17\%$ in the region $0.3<z<0.45$~\cite{Adloff:2002ey}.
All these contributions are not included in our calculation.
We will address them when comparing our results to data.

To present the numerical results, we adopt the following parameter choices.
The $c$-quark mass $m_c=1.5\gev$, and the electromagnetic coupling constant $\alpha=1/137$.
The energy of the incident lepton and proton in the laboratory frame are $E_l=27.5\gev$ and $E_p=920\gev$, respectively.
For the calculation of the $J/\psi$ production in DIS at QCD LO,
the one-loop $\alpha_s$ running and the parton distribution function (PDF) CTEQ6L1~\cite{Pumplin:2002vw} are employed,
while for the calculation up to QCD NLO, the two-loop $\alpha_s$ running and the PDF CTEQ6M~\cite{Pumplin:2002vw} are employed.
The value of $\alpha_s$ at the $Z_0$ boson mass is set to be $\alpha_s(M_Z)=0.13$.
The renormalization and factorization scales are set to be $\mu_r=\mu_f=\mu_0\equiv\sqrt{M^2+Q^2}$ as our default choice.
The $J/\psi$ wave function at the origin is determined in terms of $|R(0)|^2=0.81\gev^3$~\cite{Eichten:1995ch},
and the CS LDME can be evaluated as $\langle\mathcal{O}\rangle=9|R(0)|^2/(2\pi)\approx1.16\gev^3$.
We notice that there are many parallel extractions of the CS LDME with different strategies and data.
To be consistent with our calculation, we list here the value obtained through $J/\psi\rightarrow e^+e^-$ at QCD NLO in Reference~\cite{Braaten:2002fi},
\textit{i.e.} $\langle\mathcal{O}\rangle=(1.005\pm0.072)\gev^3$,
that extracted from the $\eta_c$ hadroproduction data, $\langle\mathcal{O}\rangle=(0.645\pm0.405)\gev^3$~\cite{Zhang:2014ybe},
and that obtained from the potential model, $\langle\mathcal{O}\rangle=1.32\gev^3$~\cite{Eichten:1995ch}.
We will find that the uncertainties in the LDMEs do not affect our phenomenological conclusions.

\begin{figure*}
\begin{center}
\includegraphics*[scale=0.55]{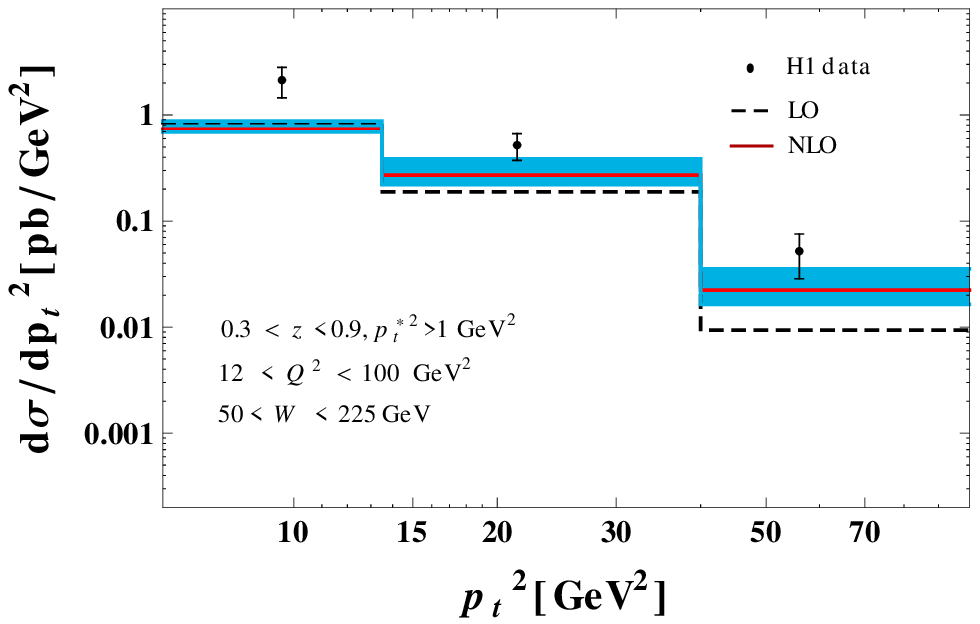}
\includegraphics*[scale=0.55]{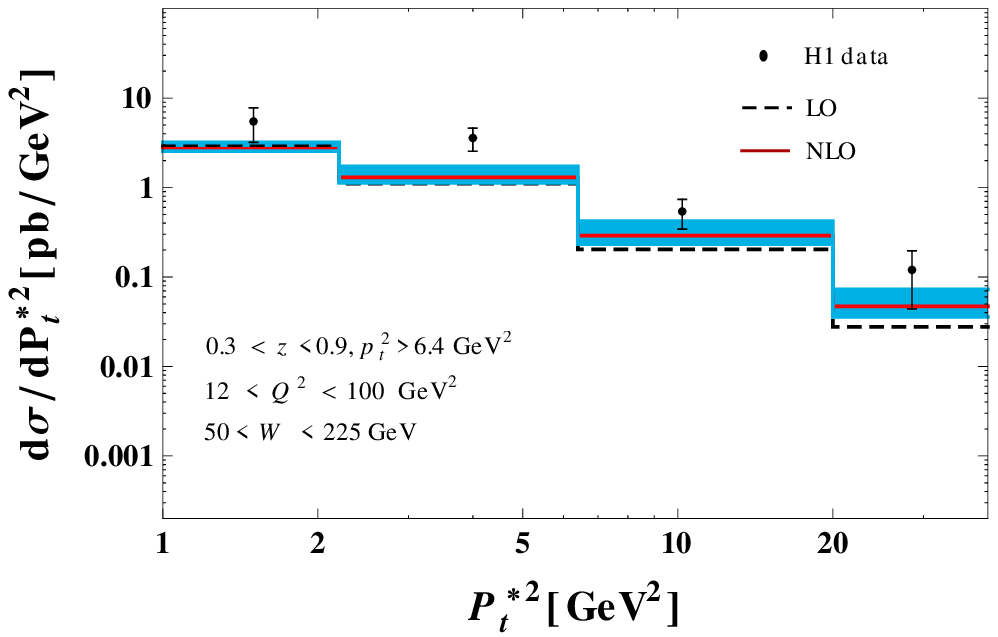}
\includegraphics*[scale=0.55]{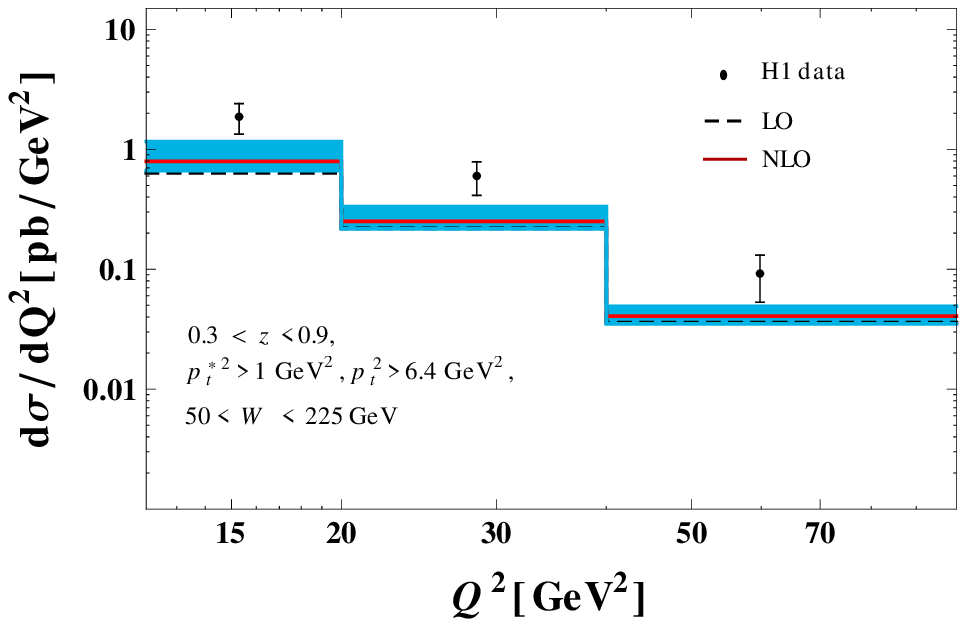}\\
\includegraphics*[scale=0.55]{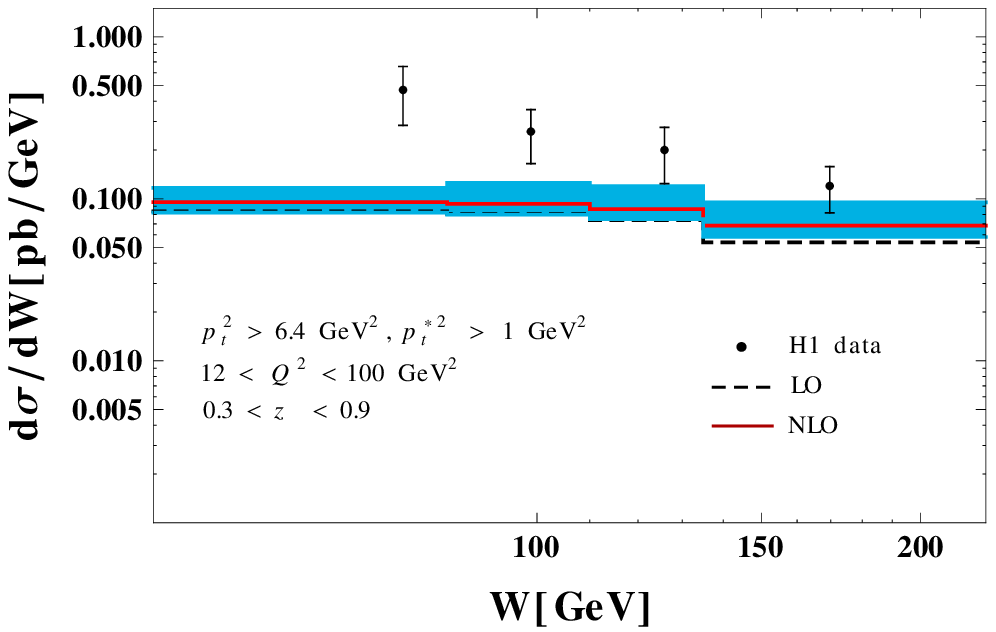}
\includegraphics*[scale=0.55]{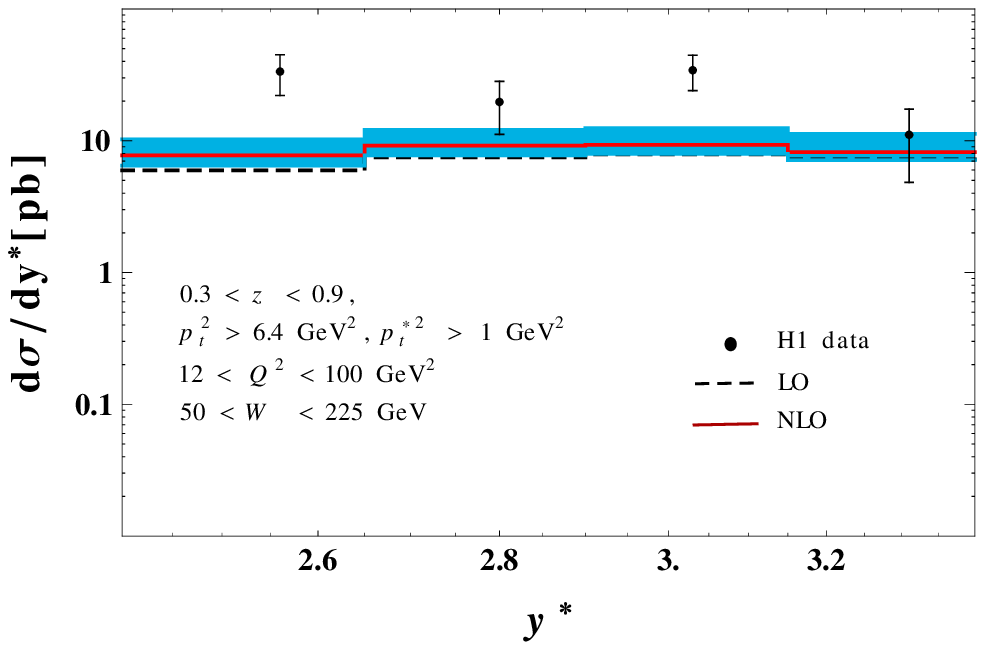}
\includegraphics*[scale=0.55]{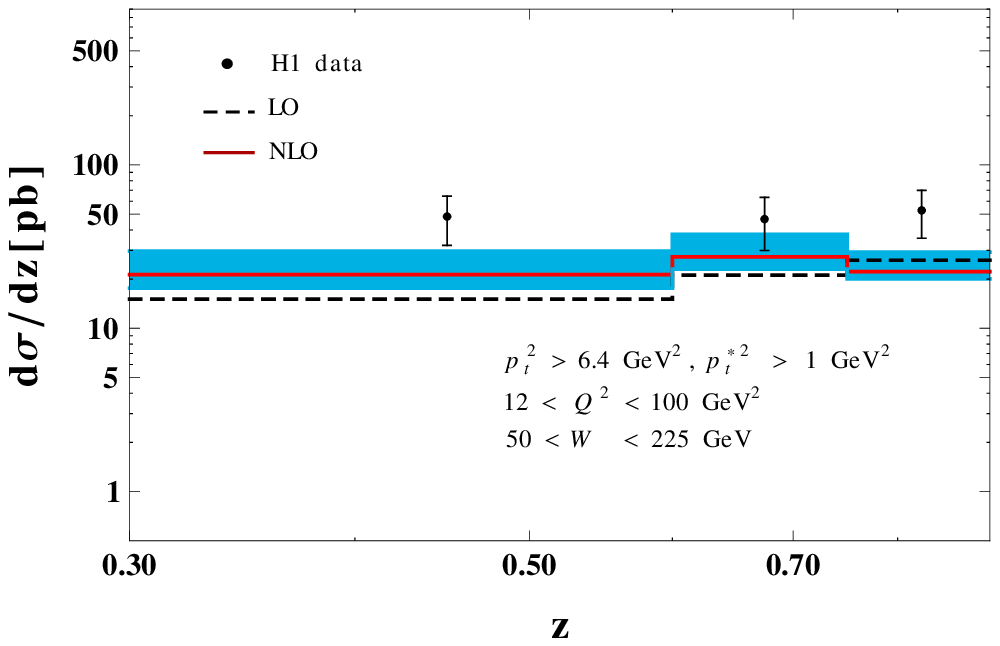}
\end{center}
\caption{
The differential cross sections of the CS $J/\psi$ production in DIS with respect to $p_t^2$, $p_t^{\star2}$, $Q^2$, $W$, $y_\psi^\star$ and $z$.
The bands are obtained by varying $\mu_r$ from $0.5\mu_0$ to $2\mu_0$.
The data are taken from Reference~\cite{Adloff:2002ey}
}
\label{fig:diffcs}
\end{figure*}

Our numerical results are presented in Figure~\ref{fig:diffcs},
including the differential cross sections with respect to $p_t^2$, $p_t^{\star2}$, $Q^2$, $W$, $y_\psi^\star$ and $z$.
In the kinematic regions we study, the $K$-factor ranges from $0.85$ to $2.38$.
As $p_t$ increases, the NLO corrections become more significant.
In the largest $p_t^2$ bin, $40\gev^2<p_t^2<100\gev^2$, the $K$-factor reaches its maximum value, $2.38$.
For the $Q^2$ bins, $12\gev^2<Q^2<20\gev^2$, $20\gev^2<Q^2<40\gev^2$ and $40\gev^2<Q^2<100\gev^2$,
the $K$-factors are $1.25$, $1.14$ and $1.13$, respectively.
As we expected, the convergence of the perturbative expansion becomes better as $Q^2$ increases.
One can easily see from the plots that, when $p_t$ and $p_t^\star$ is not too large,
the $K$-factor is quite close to 1, which indicates good convergence of the perturbative expansion in $\alpha_s$.
To study the uncertainties brought in by the uncertainties of the LDMEs,
we present the bands in Figure~\ref{fig:diffcs} covering the results for $\mu_r=0.5\mu_0$ and $\mu_r=2\mu_0$.
The upper bound of the band is generally less than 1.5 times of the central curve.

Since the $\psi(2S)$ feed down contributions are not taken into account in our calculations,
we need to estimate their magnitude while comparing our theoretical results to data.
According to our earlier discussions, the feed down contributions are about 30\% of the total $J/\psi$ production cross sections.
In contrast, the ratio of the central value of the experimental data to our NLO results range from 1.4 to 4.9,
among which, only 5 out of the 21 data points are smaller than 2.
We can conclude that even with the feed down contributions included,
the theoretical results still cannot describe the data.
At least at the NLO precision, the CO mechanism is important and necessary for understanding the $J/\psi$ production in DIS.

In summary, we studied the QCD corrections to the $J/\psi$ production in DIS.
This process is much more complicated than the $J/\psi$ photoproduction and hadroproduction ones due to the nonzero $Q^2$.
In the kinematic regions that HERA experiment concerns, we found that the $K$-factors are close to 1,
which indicates good convergence of the perturbative expansion.
With the NLO corrections included, the CS contributions are still much smaller than the experimental data.
To this end, our study iterated the importance of the color-octet mechanism.

We are indebted to Professor Geoffrey Bodwin for helpful discussions.
This work is supported by National Natural Science Foundation of China (Nos. 11405268, 11647113 and 11705034).

%

\end{document}